      [1999/12/01 v1.4c Il Nuovo Cimento]
\documentclass{cimento}

\usepackage{graphicx}
 \usepackage{mathptmx}      
\usepackage{amsmath}
\usepackage{amssymb}

\title{On compact super-massive objects without event horizon}
\author{L.~Verozub\from{ins:x}}
\instlist{\inst{ins:x}  Kharkov National University - Kharkov, Ukraine\\
E-mail: leonid.v.verozub@univer.kharkov.ua}

\PACSes{\PACSit{04.20.20.Cv}
\PACSit{04.50.+h } }
\begin{document}

\maketitle

\begin{abstract}
 This paper aims to show the possibility of the existence of super-massive compact objects with radii less than the Schwarzschild one, which is one of
the principal consequences of the author’s geodesic-invariant gravitation equations
(Ann. Phys. (Berlin), 17 (2008) 28). The physical interpretation of the solutions
of the equations is based on the conclusion that only an aggregate “space-time
geometry + used reference frame” has a physical sense.
\end{abstract}

\section{Introduction}
\label{intro}
In Einstein's theory of gravitation space-time is relative in the sense that the metric depends on the distribution of matter. However,  long before the Einstein  theory   Poincar\'{e} showed that  geometry of space  depends also on the properties of measuring instruments.  Only an aggregate  "geometry + measuring instruments" has a physical meaning, verifiable by  experience. After   Minkowski  this can be also said  about  geometry of space-time.
Some results of the attempt to actualize these ideas, and a generalization of  the vacuum Einstein's  gravitation  equations are considered in  \cite{Verozub08a}.
Such approach allows to consider gravitation both as a field in flat space-time  and as a space-time curvature.
The equations do not contradict the existing observations data. However, the physical consequences resulting from them are radically  different from the ones of general relativity at  distances of the order of the Schwarzschild radius or less than that from a dot mass.  It is very important that this fact provides a natural explanation of  modern data of   the Universe expansion.
 However, this fact  leads also to another  important physical consequence, which still has no confirmation.

Observations  give evidences for the existence of supermassive compact  cold objects in  galactic centers \cite{Genzel}.  Standard  conditions of the equilibrium of selfgravitating degenerate Fermi-gas  forbid the  existence of very massive objects. For this reason  they are usually identified with black holes.  However lack of the evidence  of the existence of 
an event  horizon  admits  also other explanations of the nature of such objects.

In \cite{Verozub96} the possibility of the  existence of supermassive  equilibrium  configurations of the degenerate  Fermi gas with radii less than the Schwarzschild one  has been considered.
Such objects  have no event horizon  and are an alternative to the hypothesis of the existence of black holes.
 It is later,  in \cite{Verozub06a},  some observable consequences of  the existence of such object in the  Galaxy center have been considered. 

However, it is seems that  a  theoretical justification of  the existence of such objects  is insufficiently convincing for observers as  it is based on conclusions resulting from  the author's  gravitation equations considered  only in a brief note  \cite{Verozub91}. In the present paper, being based on the recent paper \cite{Verozub08a} 
a simple  and a clear justification  of  possibility  of the existence of such objects is given .

\section{Gravitation equations}
\label{sec:1}
Unlike electrodynamics, Einstein's equations are not invariant with respect to
a wide class of transformations of field variables in use 
( Christoffel symbols or  metric tensor) leaving the equation of motion of test particles (geodesic
lines) invariant \cite{Petrov}. 
 For example,  all
Christoffel symbols  $\overline{\Gamma}_{\beta\gamma}^{\alpha}(x) $
obtained by the transformations
\begin{equation}
\overline{\Gamma}_{\beta\gamma}^{\alpha}(x)=\Gamma_{\beta\gamma}^{\alpha
}(x)+\delta_{\beta}^{\alpha}\ \phi_{\gamma}(x)+\delta_{\gamma}^{\alpha}%
\phi_{\beta}(x),
\label{TransformChristoffels}%
\end{equation}
where $\phi_{\beta}(x)$ are an arbitrary differentiable vector-function,
describe the same gravitational field because the geodesic equations%
\begin{equation}
\ddot{x}^{\alpha}+(\Gamma_{\beta\gamma}^{\alpha}-c^{-1}\Gamma_{\beta\gamma
}^{0}\dot{x}^{\alpha})\dot{x}^{\beta}\dot{x}^{\gamma}=0\label{GeodeicEqs}%
\end{equation}
remain invariant under transformation (\ref{TransformChristoffels}) in any
given coordinate system. (The points denote differentiation with respect to
$t=x^{0}/c$, $c$ is speed of light ). 
However, Einstein's equations are not invariant under such transformations  
because the Ricci tensor is transformed under  (\ref{TransformChristoffels}) as 
follows
\begin{equation}
  \overline{R}_{\alpha\beta}=R_{\alpha\beta}-\phi_{\alpha\beta},
\end{equation}
where $\phi_{\alpha\beta}=\phi_{\alpha;\beta}-\phi_{\alpha}\phi_{\beta}$, and
$\phi_{\alpha;\beta}$ is a covariant derivative with respect to   $x^{\alpha}$.

Each transformation (\ref{TransformChristoffels}) 
induces some mapping $g_{\alpha\beta} \rightarrow \overline{g}_{\alpha\beta}$ of the metric tensor $g_{\alpha\beta}$ which 
can be obtained by solving  some partial differential equation \cite{Eisenhart}, \cite{Verozub08a}.
Such mappings leave the equations of motion of test particles invariant, and, consequently, all metric tensors resulting from a given $g_{\alpha\beta}$ by a geodesic transformation describe the same gravitational field.
Thus, such transformations are gauge transformations of the tensor $g_{\alpha\beta}$. 
For this reason,  one can suppose that in any gravitational theory, based on  Einstein's hypothesis of the motion of test particles along geodesic lines,  only geodesic-invariant objects can have a physical sense.  

The simplest geodesic-invariant tensor  $B^{\alpha}_{\beta\gamma}$ can be formed  as follows \cite{Verozub08a}:
\begin{equation}
B_{\beta\gamma}^{\alpha}=\Pi_{\beta\gamma}^{\alpha}-\overset{\circ}{\Pi
}_{\beta\gamma}^{\alpha}%
\end{equation}
where
\begin{equation}
\Pi_{\alpha\beta}^{\gamma}=\Gamma_{\alpha\beta}^{\gamma}-(n+1)^{-1}\left[
\delta_{\alpha}^{\gamma}\Gamma_{\beta}+\delta_{\beta}^{\gamma}\Gamma_{\alpha
}\right]  
\end{equation}
and
\begin{equation}
\overset{\circ}{\Pi}_{\alpha\beta}^{\gamma}=\overset{\circ}{\Gamma}%
_{\alpha\beta}^{\gamma}-(n+1)^{-1}\left[  \delta_{\alpha}^{\gamma}%
\overset{\circ}{\Gamma}_{\beta}+\delta_{\beta}^{\gamma}\overset{\circ}{\Gamma
}_{\alpha}\right]
\end{equation}
are the Thomas symbols   for the Riemannian
space-time $V_{4}$ and for the Minkowski space-time $E_{4}$
 in a used coordinate system, respectively.
The Thomas symbols are formed
by the Christoffel symbols $\Gamma_{\alpha\beta}^{\gamma}$ and $\overset
{\circ}{\Gamma}_{\alpha\beta}^{\beta}$ of $V_{4}$ and $E_{4}$, respectively,
$\Gamma_{\alpha}=\Gamma_{\alpha\beta}^{\beta}$, and $\overset{\circ}{\Gamma
}_{\alpha}=\overset{\circ}{\Gamma}_{\alpha\beta}^{\beta}$.

In paper \cite{Verozub08a} a theory in which geodesic mappings play a role
of gauge transformations is considered.  The investigation of the problem leads
to the conclusion that  is of interest to explore physical consequences
from the following bimetric geodesic-invariant generalization of Einstein's
vacuum equations \footnote{Greek indexes run from $0$ to $3$} :
\begin{equation}
\nabla_{\alpha}B_{\beta\gamma}^{\alpha}-B_{\beta\delta}^{\epsilon}%
B_{\epsilon\gamma}^{\delta}=0.
\label{MyVacuumEqs}%
\end{equation}
The symbol $\nabla_{\alpha}$ denotes a covariant derivative in $E_{4}$ with respect to $x^{\alpha}$.



The derivation of basic equations, they physical interpretations, and a correct utilization are possible only owing to  reconsider of a deep problem of space-time relativity with respect to measuring instruments, going back to Poincar\'{e} fundamental ideas. According to \cite{Verozub08a}  both the space-time,  $E_{4}$ and $V_{4}$, have physical meaning since only
 aggregate ``space-time geometry + frame of reference'' have a physical meaning.
We can consider space-time both as the Minkowski one in an inertial reference frames \footnote{By inertial reference frame we mean the  frame  in which Newton's  first and second laws are obeyed globally.}
, and  as the Riemannian space-time   with curvature other than zero -- in  the so-called proper reference frames of the field, the body of reference of which is formed by  particles moving in this gravitational field.  For purpose of this paper  it is essentially only the important conclusion that an observer, which explores gravitational field of a distant massive compact object in an inertial reference frame, can consider  global space-time  as the Minkowski one.

In order to see better the relationship between (\ref{MyVacuumEqs}) and the
Einstein vacuum equation, it should  be noted that we can select a gauge
condition for  the Christoffel symbols as follows:
\begin{equation}
Q_{\alpha}=\Gamma_{\alpha\beta}^{\beta}-\overset{\circ}{\Gamma^{\beta}%
}_{\alpha\beta}=0.\label{AdditionalConditions}%
\end{equation}
At such covariant gauge condition  eq. (\ref{MyVacuumEqs}) coincides  with the vacuum Einstein's equations.
( At least locally).
Therefore, 
the observer which is located in an inertial reference frame far away from a  compact object can describe  gravity of the object field
 by the spherically-symmetric solution of  the vacuum
Einstein equations in the Minkowski  space-time (in which $g_{\alpha\beta}(x)$
is simply a tensor field) at the additional condition $Q_{\alpha}=0$ . 
Such equations somewhat resemble the equations of classical electrodynamics for 4-potential together with some specific gauge  conditions.  
It is that  will be used in the next section.

\section{Gravitational Energy of a matter  sphere }
\label{sec:2}
If a matter sphere is considered as formed by means of consecutive injections 
 of thin spherically-symmetric layers of the matter from infinity   \cite{Landau}, an equation for finding of the
 gravitational energy of the sphere can be found by using
the  vacuum solution of the gravitational equations  \cite{Verozub08a}
for a dot source. 

The differential equations of the motion of a test particle  in the
spherically-symmetric gravitational field of a dot mass $M$ from the point of view of a distant observer 
can be found from the Lagrangian 
\begin{equation}
L=-m\,c\,[A\dot{r}^{2}+B(\dot{\theta}^{2}+\sin^{2}\theta\ \dot{\varphi}%
^{2})-c^{2}C]^{1/2} \label{LagrTestParticls},%
\end{equation}
where  $(t,r,\varphi)$ are the spherical coordinates,
 $m$ is the particle mass, 
$A$, $B$ and $C$ are the functions of the radial coordinate $r$.

The equations of the motion of a test particle in the field of a point mass $M$, based on the 
 solution  of the Einstein equations under the condition $Q_{\alpha}=0$ in Minkowski space-time, are of the form  \cite{Verozub08a}

\begin{equation}
{\dot{r}}^{2}=(c^{2}C/A)[1-(C/\overline{E}^{2})(1+r_{g}^{2}\overline{J}%
^{2}/f^{2})], , \label{EqsMotionTestPart1}%
\end{equation}%
\begin{equation}
\dot{\varphi}=c\;C\overline{J}/r_{g} f^{2}\overline{E} ,
\label{EqsMotionTestPart2}%
\end{equation}
where  $f=(r_{g}^{3}%
+r^{3})^{1/3}$, $C=1-rg/f$ , $\dot{r}=dr/dt$, $\dot{\varphi}=d\varphi/dt$ ,
$\overline{E}=E/mc^{2}$, $\overline{J}=J/r_{g}mc$, $E$ and $J$ are the energy
and angular momentum of the particle, $r_{g}=2 G M/c^{2}$ is the Schwarzschild radius of the central object.,
$G$ is the gravitational constant.

If the radial distance $r$ from the dot massive object is many larger than the
Schwarzschild radius $r_{g}$ of the object , physical consequences following
from  (\ref{MyVacuumEqs}) are very close to the ones resulting from
Einstein's equations. However, they are very different when $r$ is of the
order of $r_{g}$, or less than that. 

From point of view of a distant observer  free-falling particles move up to the center.
The  spherically-symmetric solution has no event horizon  \cite{Verozub08a}, and this fact  is not a consequence of some specific coordinate system. 

It follows from  eq. (\ref{EqsMotionTestPart1})  that at  $J=0$ the energy of a rest particle at the distance $r$ from the center is :
\begin{equation}
E=mc^{2}\sqrt{C}.
\end{equation}
At the condition $\overset{\_}{r}=r/r_{g}\ll1,$ it yields
\begin{equation}
E\simeq mc^{2}-\frac{GMm}{r}.
\end{equation}
If $\overset{\_}{r}=r/r_{g}\gg1,$we obtain at first order to $\overset{\_}%
{r}^{-1}:$%
\begin{equation}
E\approx  \frac{mc^{2}r^{3/2}}{2\sqrt{6}G^{3/2}M^{3/2}}.
\end{equation}
In this case the gravitational energy of the test particle tends to zero when $r\rightarrow 0$.

Consequently, the difference between  the gravitational energy of the  thin matter layer of mass $\delta m$ at the distance $r$ from the center and 
the one at infinity is  
 $\delta m\, c^{2}\sqrt{C}-\delta m\, c^{2}$. The
gravitational energy of a material sphere, formed by means of consecutive
additions of thin spherically-symmetric layers of the matter \cite{Landau}
is equal to
\begin{equation}
\mathcal{E}_{g}=-\int(1-\sqrt{C})dm,\label{Egr_integral}%
\end{equation}
where $dm=4\pi\rho r^{2}dr,$and $\rho$ is the mass-energy density.

If the matter sphere is an  isentropic ideal fluid in equilibrium state, 
(\ref{Egr_integral}) can be transformed to more useful form.  The equilibrium
condition of the ideal isentropic fluid is of the form \cite{Verozub08b}
\begin{equation}
(\varkappa^{2}C)^{\prime}=0,
\end{equation}
where prime denotes differentiation with respect to $r$,  $\varkappa=w/\rho
_{0}c,$, $\rho_{0}\approx m_{n} n$ is matter rest-density, where $m_{n}$ is the neutron mass and $n$ is the particles number density, 
$w$ is the enthalpy per unit
volume. Therefore, we can use the following equation of the sphere
equilibrium:%
\begin{equation}
\varkappa^{2}C=C_{0},\label{EquilibriumEquationSphere}%
\end{equation}
where
\begin{equation}
\varkappa=1+\frac{P}{\rho_{0}c^{2}},
\end{equation}
$P$  is the  presure,%
\begin{equation}
C_{0}=1-\frac{r_{g}}{^{(r_{g}^{3}+R^{3})^{1/3}}},
\end{equation}
where $r_{g}=2GM/R$ is the Schwarzschild radius of the sphere. 

By using (\ref{EquilibriumEquationSphere}),  eq. (\ref{Egr_integral}) can be
written as follows:%
\begin{equation}
\mathcal{E}_{g}=-Mc^{2}+c^{2}\sqrt{C_{0}}\int\varkappa^{-1}dm.
\end{equation}
For estimates we can use it in the more simple form%
\begin{equation}
\mathcal{E}_{g}=-Mc^{2}+Mc^{2}\overset{\_}{\varkappa}^{-1}\sqrt{C_{0}},
\end{equation}
where $\overset{\_}{\varkappa}$ is an averaged value of $\varkappa$ over the sphere.

As $r_{g}/R\ll1$б   the functions $C_{0}\approx  1-  r_{g}/R$, and  $\varkappa^{-1} \approx 1-P/\rho_{0} c^{2}$  because usually $P/\rho_{0}c^{2}\ll1$ 
By using a polytropic equation of the fluid state we obtain  that
 the gravitational energy of
the object with  radius $R\gg r_{g}$ is
\begin{equation}
\mathcal{E}_{g}=-\frac{GM^{2}}{R}-\int PdV,
\end{equation}
where $dV$ is a volume element. The gravitational energy is very small as
compared with full energy $Mc^{2}$ of the object

For the  object with  radius $R<r_{g}$ the situation is another. If
$R/r_{g}\ll1,$ the value of $C_{0}$ is close to zero, and the gravitational
energy is close to $-Mc^{2}.$ 

\section{Minimization of total energy}
\label{sec:3}
For the total energy of the uniform sphere we obtain the
relationship
\begin{equation}
\mathcal{E}=\mathcal{E}_{0}+\mathcal{E}_{k}+\mathcal{E}_{g}=\mathcal{E}%
_{0}+\mathcal{E}_{k}-\mathcal{E}(1-\overset{\_}{\varkappa}^{-1}\sqrt{C_{0}}),
\end{equation}
where $\mathcal{E}_{0}=Mc^{2},\mathcal{\ }$ $\mathcal{E}_{k}$ is the internal
energy of the fluid. Thus,
\begin{equation}
\mathcal{E=}\frac{\mathcal{E}_{0}+\mathcal{E}_{k}}{2-\overset{\_}{\varkappa
}^{-1}\sqrt{C_{0}}},\label{FinalEnergyEquation}.
\end{equation}
The magnitude $C_{0}$ contains the total mass of the object
\begin{equation}
M=M_{0}+M_{k}+M_{g},
\end{equation}
where $M_{0}\approx m_{n}c^{2},$ $M_{k}=\mathcal{E}_{k}/c^{2},$ $M_{g}%
=\mathcal{E}_{g}/c^{2}$. If $r_{g}/R \ll1,$
$M\approx M_{0}+M_{k}$. 
 If $r_{g}/R\gg1$, then 
 $M\approx(M_{0}%
+M_{k})/2.$ For this reason in  (\ref{FinalEnergyEquation}) we can replace approximately $M$ in
$C_{0}$  by      $M=\lambda(M_{0}+M_{k}),$ where $\lambda$ is a parameter which lays in
the range $1\div1/2$.

 For an uniform sphere of a nonrelativistic
degenerate electronic or neutron ideal gas
\begin{equation}
\mathcal{E}_{k} \propto   \frac{N^{5/3}}{R^{2}},
\end{equation}
where $N$ is the  total number of  particles, and $R$ is the radius of the
sphere. Therefore, $\mathcal{E}$ can be considered as the function of $N$
 and $R$.

 Figures 1-4 show  the total energy  of  uniform spherical configurations  of degenerate Fermi-gas, considered as  an 
ideal isentropic  fluid,   as a function of $R$ for the several  values of $N$.
It is easy to verify that the parameter $\lambda$  affect scarcely the plots.

\begin{figure}[h]
\includegraphics[width=6cm,height=4cm]{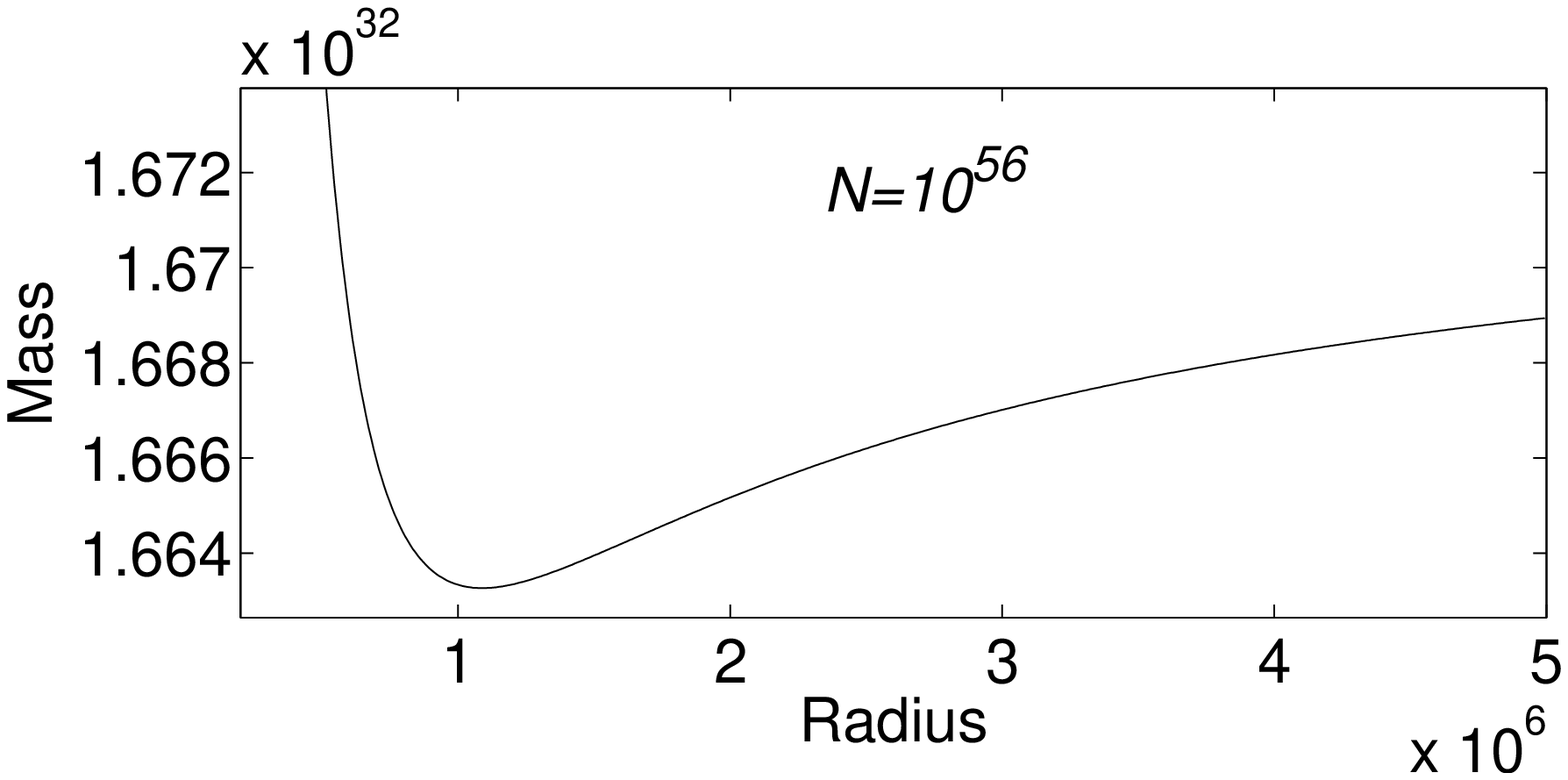}
\hfill
\includegraphics[width=6cm,height=4cm]{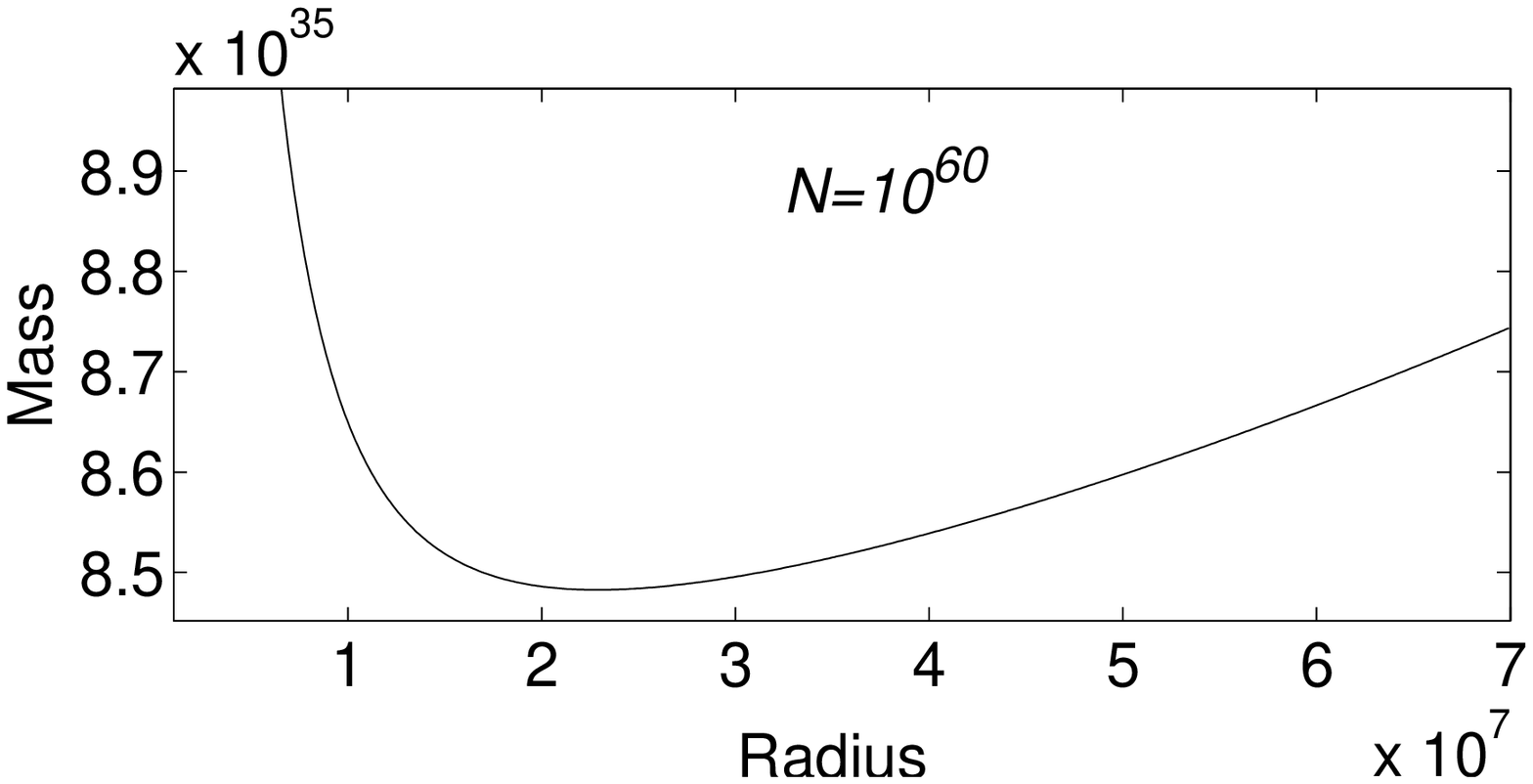}\newline%
\hfill
\includegraphics[width=6cm,height=4cm]{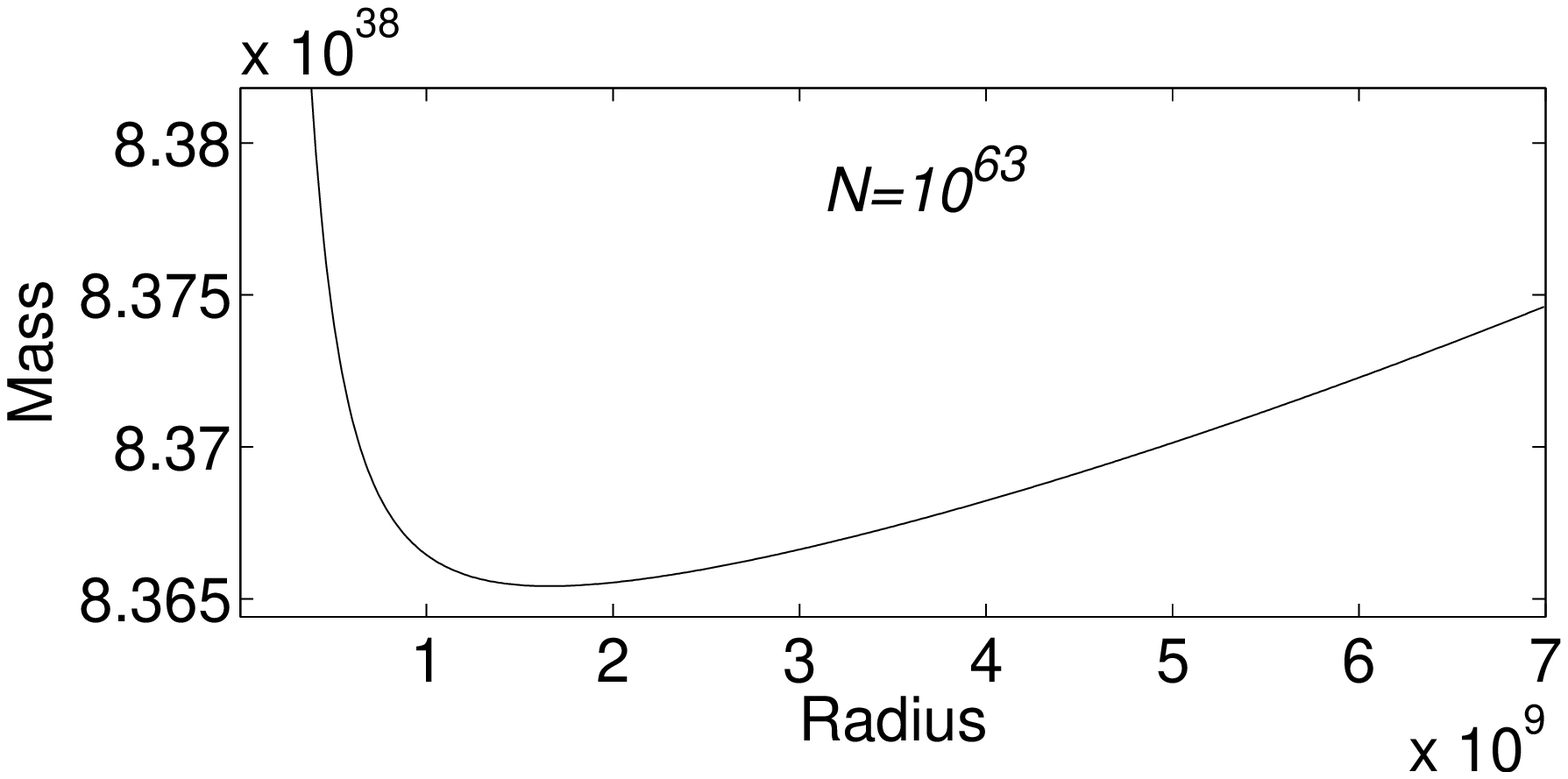}
\hfill
\includegraphics[width=6cm,height=4cm]{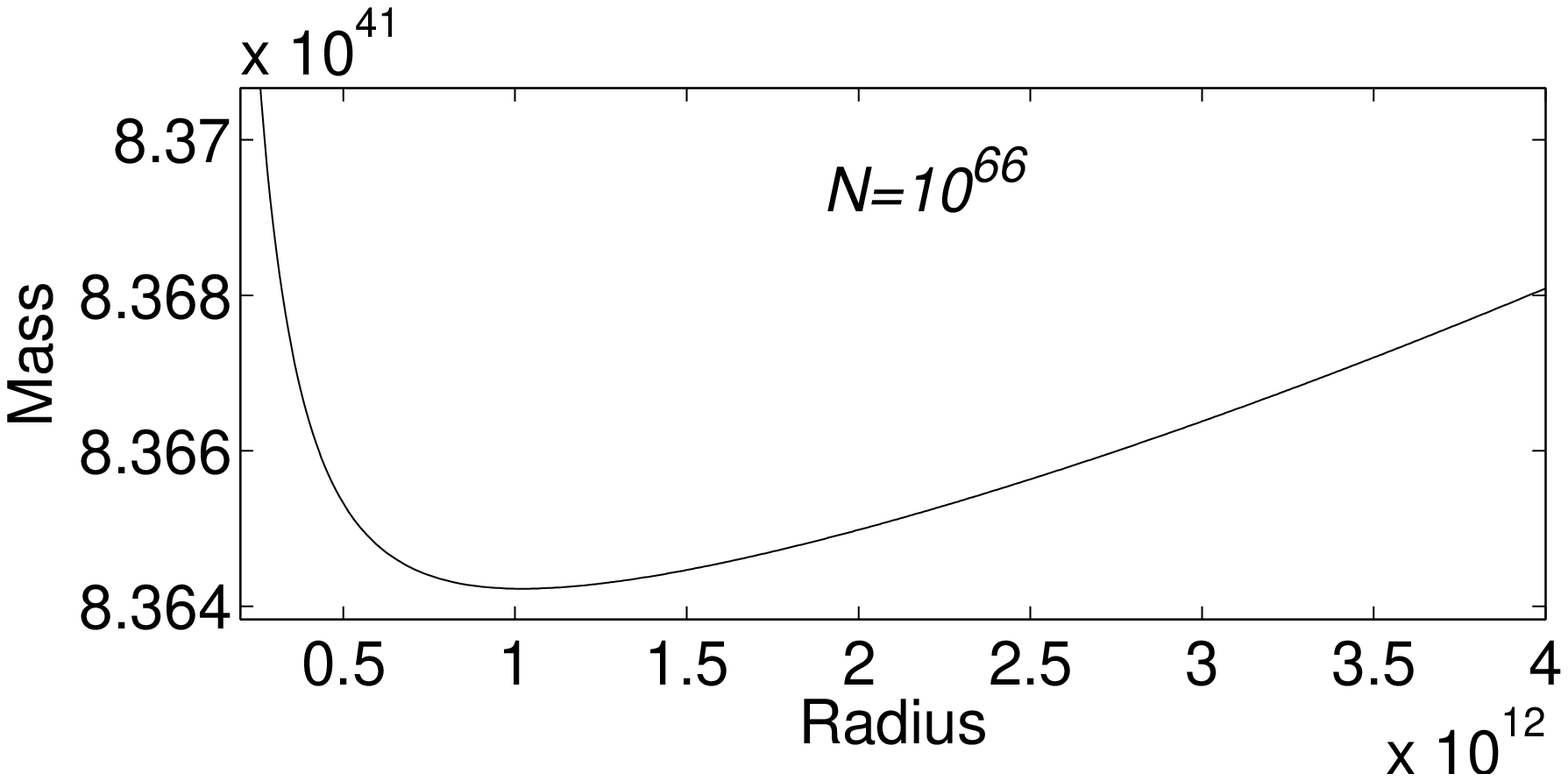}\newline%
\caption{The mass $M=\mathcal{E}/c^{2}$ of the configuration as a function of total number of particles .}%
\label{fig:erptsqfit}%
\end{figure}

It follows from the figures  that  the  total energy has  the  minimum not only at $N=10^{56}$ (
neutron star), but also at many larger values of $N$. They are
supermassive objects with radiuses $R$ larger of the Schwarzschild radius.   Table 1 shows 
the masses $M$, radius $R$, average density $\rho_{0}$, and the Schwarzschild radiuses $r_{g}$ of the stable configurations determined by the plots minimum. The first three configuration are polytropic neutron configurations, and the configuration at last line is the electronic one. According to the used  simple  model the maximal mass  of the purely neutronic configurations  is  approximately $M=10^{5}$ of  the Sun mass.  However,  electronic configurations are  possible up to $M=10^{12}$ of  the Sun mass..

\begin{table}[b]
\caption{Features of the stable configurations determined by the minimum of the total energy .}
\label{tab:3}
\begin{tabular}{@{}lllll@{}}
\hline
  $N$ & $ M$ & $ R$  & $\rho_{0}$ & $r_{g}$\\
\hline
$10^{56} $ & $1.66 \times 10^{32}$ & $ 1.11 \times10^{6}$  &  $ 2.92 \times 10^{13}$  & $2.46 \times 10^{4} $ \\
$10^{60} $ & $8.50 \times 10^{35}$ & $ 2.57 \times10^{7}$  &  $ 2.35 \times 10^{13}$  & $1.26 \times 10^{8} $ \\
$10^{63} $ & $8.36 \times 10^{38}$ & $ 1.86 \times10^{9}$  &  $ 6.20 \times 10^{10}$  & $1.24 \times 10^{11} $ \\
$10^{66} $ & $8.36 \times 10^{41}$ & $ 1.14 \times10^{12}$  &  $ 2.69 \times 10^{5}$  & $1.24 \times 10^{14} $ \\
\hline
\end{tabular}
\end{table}
\section{Conclusion}
Despite the fact that the above calculations  yield  rather qualitative than quantitative results, they show clearly that 
the according to equations of gravitation  (\ref{MyVacuumEqs}), which do not contradict available observation data,  there are  stable supermassive configurations of degenerate  Fermi-gas with radiuses less than the Schwarzschild radius. Just such object can be located in the Galaxy center  \cite{Verozub06a}.

\end{document}